\documentclass[12pt]{article}
\usepackage{graphicx}
\begin{document}
\centerline {\bf Monte Carlo simulations of Sznajd models}

\bigskip
\centerline{Dietrich Stauffer}

\centerline{Institute for Theoretical Physics, Cologne University}

\centerline{D-50923 K\"oln, Euroland}

\bigskip
{\tt www.thp.uni-koeln.de; stauffer@thp.uni-koeln.de}

\bigskip
{\small The Sznajd model in less than a year has found several followers. 
An isolated person does not convince others; a group of people
sharing the same opinions influences the neighbours much more easily.
Thus on a square lattice, with variables +1 (Democrats) and --1
(Republicans)
on every lattice site, a pair (or plaquette) of neighbours convinces its
six (eight) nearest neighbours of its own opinion if and only if all
members
of the pair (plaquette) share the same opinion. The generalization to many
possible states is used to explain the distribution of votes among
candidates in Brazilian local elections.}
\bigskip


Review; published in Journal of Artificial Societies and Social Simulation 5, 
No. 1, paper 4 (2002), available only electronically: jasss.soc.surrey.ac.uk
\end{document}